\documentclass{article}

\usepackage{PRIMEarxiv}

\usepackage[utf8]{inputenc} %
\usepackage[T1]{fontenc}    %
\usepackage{hyperref}       %
\usepackage{url}            %
\usepackage{booktabs}       %
\usepackage{amsfonts}       %
\usepackage{nicefrac}       %
\usepackage{microtype}      %
\usepackage{lipsum}
\usepackage{fancyhdr}       %
\usepackage{graphicx}       %
\usepackage{subcaption}
\usepackage{caption}       %
\graphicspath{{figures/}}     %
\usepackage{epstopdf}
\usepackage[table,xcdraw]{xcolor}
\usepackage[table]{xcolor}  %
\usepackage{amsmath}  %

\usepackage{multirow}
\usepackage{placeins}
\usepackage{soul,color}
\title{MD-HIT: Machine learning for materials property prediction with dataset redundancy control
\thanks{\textit{\underline{Citation}}: 
\textbf{Qin Li et al.. MD-HIT. 12 Pages.... DOI:000000/11111.}} 
}

\author{
 Qin Li \\
 College of Big data and statistics\\
  Guizhou University of Finance and Economics\\
  Guiyang, China 550050 \\
  \And
  Nihang Fu\\
 Department of Computer Science and Engineering\\
  University of South Carolina\\
  Columbia, SC 29201 \\  
   \And
 Sadman Sadeed Omee\\
 Department of Computer Science and Engineering\\
  University of South Carolina\\
  Columbia, SC 29201 \\  
   \And
 Jianjun Hu *\\
 Department of Computer Science and Engineering\\
  University of South Carolina\\
  Columbia, SC 29201 \\
  \texttt{jianjunh@cse.sc.edu} \\
}

\begin{document}
\maketitle

\begin{abstract}

Materials datasets are usually featured by the existence of many redundant (highly similar) materials due to the tinkering material design practice over the history of materials research. For example, the materials project database has many perovskite cubic structure materials similar to SrTiO$_3$. This sample redundancy within the dataset makes the random splitting of machine learning model evaluation to fail so that the ML models tend to achieve over-estimated predictive performance which is misleading for the materials science community. This issue is well known in the field of bioinformatics for protein function prediction, in which a redundancy reduction procedure (CD-Hit \cite{li2006cd}) is always applied to reduce the sample redundancy by ensuring no pair of samples has a sequence similarity greater than a given threshold. This paper surveys the overestimated ML performance in the literature for both composition based and structure based material property prediction. We then propose a material dataset redundancy reduction algorithm called MD-HIT and evaluate it with several composition and structure based distance threshold sfor reducing data set sample redundancy. We show that with this control, the predicted performance tends to better reflect their true prediction capability. Our MD-hit code can be freely accessed at \url{https://github.com/usccolumbia/MD-HIT}

\end{abstract}


\section{Introduction}

\begin{figure}[ht]
  \centering

      \begin{minipage}[c]{0.499\textwidth}
        \centering
        \includegraphics[width=\textwidth]{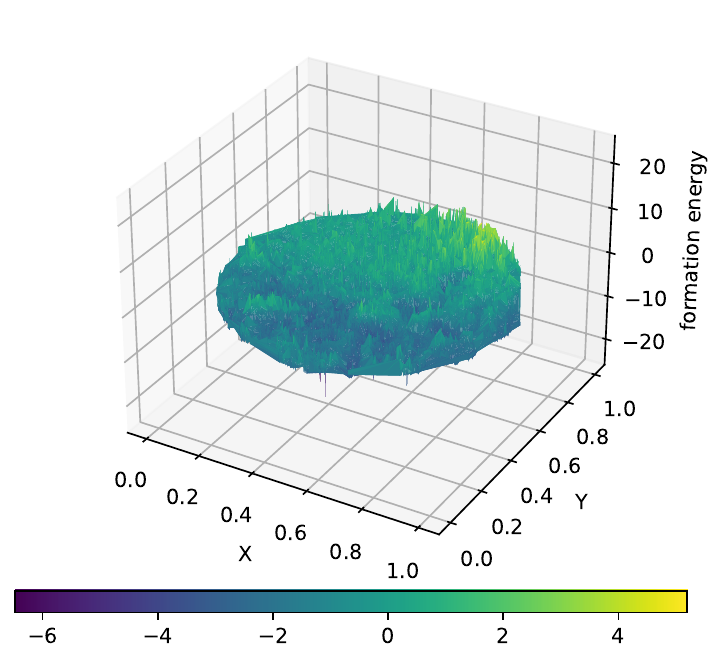}
        \subcaption{formation energy landscape}
    \end{minipage}
    \begin{minipage}[c]{0.496\textwidth}
        \centering
        \includegraphics[width=\textwidth]{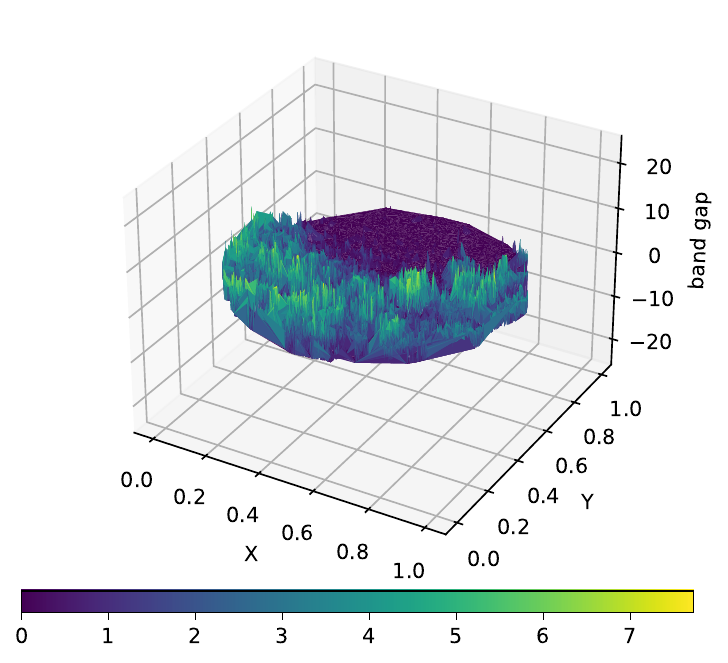}
        \subcaption{band gap landscape}
    \end{minipage}\\

  \caption{Landscape of material properties. In many continuous landscape areas, there exist crowded samples with similarity properties, which makes it trivial to predict the property if a query sample is located in these areas with multiple neighbors in the training set.}
  \label{fig:landscape}
\end{figure}

Density functional theory (DFT) level accuracy of material property prediction \cite{xie2018crystal} and $>$0.95 $R^2$ for thermal conductivity prediction \cite{chen2019machine} with less than a hundred training samples have been routinely reported recently by an increasing list of machine learning algorithms in the material informatics community. In \cite{jha2022moving}, an AI model was shown to be able to predict formation energy of a hold-out test set containing 137 entries from their structure and composition with a mean absolute error (MAE) of 0.064 eV/atom which significantly outperform the performance of DFT computations for the same task (discrepancies of >0.076 eV/atom). In another related work in Nature Communication by the same group \cite{jha2019enhancing}, a mean absolute error (MAE) of 0.07 eV/atom was achieved for composition only based formation energy prediction using deep transfer learning, which is comparable to the MAE of DFT-computation. Pasini et al \cite{pasini2020fast} reported that their multitasking neural networks can estimate the material properties (total energy, charge density and magnetic moment) for a specific configuration hundreds of times faster than first-principles DFT calculations while achieving comparable accuracy. In \cite{chen2019graph}, the authors claimed their graph neural network models can predict the formation energies, band gaps, and elastic moduli of crystals with better than DFT accuracy over a much larger data set. In \cite{faber2017prediction}, Farb et al. showed numerical evidence that ML model predictions deviate from DFT less than DFT deviates from experiment for all nine properties that they evaluated over the QM9 molecule dataset. They also claimed the out-of-sample prediction errors with respect to hybrid DFT reference were on par with, or close to, chemical accuracy. In \cite{tian2022information}, Tian et al reported that current ML models can achieve  accurate property-prediction (formation energy, band gap, bulk and shear moduli) using composition alone without using structure information, especially for for compounds close to the thermodynamic convex hull. However, this good performance may be partially due to the over-represented redundancy in their test samples obtained with 6:2:2 random selection from matminer datasets without redundancy control. To illustrate this point, Figure \ref{fig:landscape} shows the formation energy and band gap landscape over the MP composition space, which is generated by mapping the Magpie features of all MP unique compositions to the 2D space using t-SNE and then plot the surface. Both figures show that there exist a large number of local areas with smooth or similar property values. Random splitting of samples in those areas into training and test sets may lead to information leakage and over-estimation of the prediction performance.

Despite these encouraging successes, the DFT accuracy reports of these ML models for material property prediction should be cautiously interpreted as they are all average performance evaluated over mostly randomly held-out samples that come from unexpectedly highly redundant datasets. Materials databases such as Material Project and OQMD are characterized by the existence of many redundant (highly similar) materials due to the tinkering material design practice over the history of materials research. For example, the materials project database has many perovskite cubic structure materials similar to SrTiO$_3$. This sample redundancy within the dataset makes the random splitting of machine learning model evaluation to fail so that the ML models tend to achieve over-estimated predictive performance which is misleading for the materials science community. This issue is well known in the area of ecology \cite{roberts2017cross} and bioinformatics for protein function prediction, in which a redundancy reduction procedure (CD-Hit \cite{li2006cd}) is required to reduce the sample redundancy by ensuring no pair of samples has a sequence similarity greater than a given threshold e.g. 95\% sequence identity. In a recent work in 2023, it was also shown that excellent benchmark score may not imply good generalization performance \cite{li2023critical}.

The over-estimation of the ML performance for materials has been investigated in a few studies. In \cite{meredig2018can}, Meredig et al. examined extrapolation performance of ML methods for materials discovery. They found that traditional ML metrics (even with cross-validation (CV)) overestimate model performance for materials discovery and introduce the leave-one-(material) cluster-out cross-validation (LOCO CV) to objectively evaluate the extrapolation performance of ML models. They especially highlighted that materials scientists often intend to extrapolate with trained ML models, rather than interpolate to find new functional materials and sampling in materials training data is typically highly non-uniform. So the high interpolation performance of ML models trained with datasets with high sample redundancy (e.g. due to doping) does not indicate their strong capability to discovery new materials (out of dotmain (OOD) samples). They showed that current ML models have much higher difficulty to generalize from the training clusters to a distinct test cluster. They suggested the use of uncertainty quantification (UQ) on top of ML models to evaluate and explore candidates in new regions of design space. Stanev et al. \cite{stanev2018machine} also discussed this generalization issue across different superconductor family. In \cite{xiong2020evaluating}, Xiong et al. propose K-fold forward cross-validation (FCV) as a new way for evaluating exploration performance in materials property prediction by first sorting the samples by their property values before CV splitting. They showed that current ML models' prediction performance were actually very low as shown by their proposed 
FCV evaluation method and the proposed exploratory prediction accuracy. A similar study for thermal conductivity prediction \cite{loftis2020lattice} also showed that when ML models are trained with low property values, they are usually not good at predicting samples with high property values, indicating the weak extrapolation capability. These studies show the need for the material property model developers to focus more on extrapolative prediction performance rather than average interpolation performance over test samples with high similarity to training samples due to dataset redundancy. 

The material datasets redundancy issue has also been studied recently from the point of view of training efficient ML models or achieving sample efficiency. In \cite{magar2023learning}, Magar and Farimani proposed an adaptive sampling strategy to generate/sample informative samples for training machine learning models in the lowest amounts of data. They assumed that informative samples for a model are those with the highest K(e.g. 250) MAE in the test set, which are added to the initial 1000 training set iteratively. Another selection approach is to add samples similar to data points of the train set having the maximum MAE during training. They showed that their sampling algorithms can create smaller training sets that obtain better performance than the baseline CGCNN model trained with all training samples. This approach can be used with active learning to build high performance ML models in a data efficient way. In a more recent work \cite{li2023redundancy}, Li et al. studied the redundancy in large material datasets and found that a significant degree of redundancy across multiple large datasets is present for various material properties and that up to 95\% of data can be removed from ML model training with little impact on prediction performance for test sets sampled randomly from the same distribution dataset. They further showed that the redundant data is due to over-represented material types and does not help improve the low performance on out-of-distribution samples. They proposed a pruning algorithm similar to \cite{magar2023learning} which first splits the training set into A and B and then train a ML model on A and evaluates the prediction errors on samples in B. After that the test samples with low MAEs are pruned and the remaining samples are merged and split into A and B again and so on. Both approaches rely on the iterative training of ML models and are specific to a given material property. The also proposed an uncertainty quantification based active learning method to generate sample efficient training set for model training. While these works recognize the possibility to build data-efficient training set, they did not mention the how redundancy can affect the over-estimated ML model performance commonly seen in literature. Moreover, all approaches for building informative training set are material property specific, making it difficult to generate a single non-redundant benchmark dataset for benchmarking material property prediction algorithms for all material properties. Another limitation of these methods is that they show different similarity thresholds when applied to different datasets, which makes the resulting non-redundant datasets to have different minimum distances among the samples. 

Since material property prediction research is now pivoting toward developing ML models with high accuracy, that are generalizable and transferable between different materials (including materials of different families), healthy evaluation of ML algorithms is needed to recognize the limitation of existing ML models and to invent new models with essential process. Within this context, reducing the dataset redundancy of both training set and test sets can avoid the over-estimation of the ML model performance, ameliorate the training bias towards samples in crowded areas, and push the model developers to focus on improving extrapolation performance instead of only interpolation performance.

In this paper, we argue the importance of redundancy control in the training and test set selection to achieve objective performance evaluation. Neglecting this has lead to many overestimated ML performances as reported in the literature for both composition based and structure based material property prediction. We then conduct the ML experiments to show that the over-estimated models usually fail for samples that are distant to training samples (lack of extrapolation performance). We then developed two redundancy reducing algorithms (CD-hit-composition and CD-hit-structure) with open-sourced code for reducing the dataset redundancy of both composition datasets and structure datasets. These two algorithms are based on composition and structure based distance metrics, which are used to add samples that are above a defined distance threshold. After this data redundancy control, the dataset can then be splitted randomly into training, validation, and test sets to achieve objective performance evaluation. We show that with this dataset redundancy control, the predicted performance tends to reflect their true prediction capability.

\section{Method}
\label{sec:headings}

\subsection{MD-HIT-composition algorithm for redundancy reduction of composition datasets}
The early version of CD-HIT algorithm \cite{li2006cd} of bioinformatics was originally developed to handle large-scale sequence datasets efficiently. It employs a clustering approach to group similar sequences together based on a defined sequence identity threshold. Within each cluster, only one representative sequence, called the "centroid," is retained, while the rest of the highly similar sequences are considered duplicates and removed. However, the clustering approach is still inefficient to deal with datasets with hundreds of thousands of sequences. The next generation of CD-HIT further improved the efficiency by using a greedy algorithm \cite{fu2012cd}. 
Both of our MD-HIT-composition and MD-HIT-structure redundancy reduction algorithms are designed based on this idea, which are greedy incremental algorithms. In our case, the MD-HIT starts the selection process with a seed material (default to be H2O). And then it sorts the remaining materials by the number of atoms instead of the formula lengths and then one-by-one classifies them as a redundant or representative material based on its similarities to the existing representatives already selected into the cluster. The composition similarities are estimated using the ElMD (Earth Movers' Distance) package, which provides the options to choose linear, chemically derived, and machine learned similarity measures. By default, we used the mendeleev similarity and the magpie similarity \cite{ward2016general} for our non-redundant composition dataset generation. The magpie distance function is defined as the Euclidean distance of a given set of material composition magpie feature vectors such as the widely used magpie features \cite{ward2016general}. In the matminer materials informatics package, there are several other material composition descriptors that can also be used as well. Here we focused on ElMD and the magpie feature based distance function for redundancy control of composition datasets for materials property prediction.

The complete composition similarity metrics can be found in Table \ref{tab:similarity}.

\begin{table}[th]
\centering
\caption{Composition similarity categories and metrics}
\begin{tabular}{|c|c|}
\hline
\rowcolor[HTML]{C9D9D4} 
 \textbf{Category} & \textbf{metric} \\
\hline
\multirow{4}{*}{Linear} & mendeleev \\
& petti \\
& atomic \\
& mod\_petti \\
\hline
\multirow{6}{*}{Chemically Derived} & oliynyk \\
& oliynyk\_sc \\
& jarvis \\
& jarvis\_sc \\
& magpie \\
& magpie\_sc \\
\hline
\multirow{5}{*}{Machine Learnt} & cgcnn \\
& elemnet \\
& mat2vec \\
& matscholar \\
& megnet16 \\
\hline
\end{tabular}
\label{tab:similarity}
\end{table}

\subsection{MD-HIT-Structure algorithm for redundancy reduction of structure datasets}

MD-HIT-structure algorithm uses the same greedy adding approach of the MD-HIT-composition except it uses a structure based distance metric. However, due to the varying number of atoms of different crystals, it is non-trivial to compare the similarity of two given structures given most of structure descriptors tend to have different dimension for structures of different number of atoms. Here we chose two structure distances for redundancy reduction. One is the distance metric based on XRD features calculated from crystal structures. We used a Gaussian smoothing operation to first smooth the calculated XRD using the Pymatgen XRDCalculator module and then sample 900 points even distributed between 0 and 90 degree, which leads to XRD features of a fixed 900 dimension. 

We also selected the OrbitalFieldMatrix feature to calculate the distances of two structures. This feature has also been used in \cite{magar2023learning} to select informative samples for ML model training. It is a set of descriptors that encode the electronic structure of a material. These features provide information about the distribution of electrons in different atomic orbitals within a crystal structure. These features provide a comprehensive representation of the electronic structure and bonding characteristics of materials and is of fixed dimension (1024). 

Similar to the MD-Hit-composition, MD-Hit-structure algorithm also starts the selection process with a seed material (default to be H2O) put in the non-redundant set. And then it sorts the remaining materials in the candidate set by the number of atoms instead of the formula lengths and then one-by-one classifies them as a redundant or representative material based on its similarities (we use Euclidean distance of XRD features or OFM features) to the existing representatives already selected into the non-redundant set. Redundant samples are discarded while non-redundant ones are added to the non-redundant set until the candidate set is empty.

\subsection{Composition based materials property prediction algorithms}

We evaluate two state-of-the-art composition based material property prediction algorithms including Roost \cite{goodall2020predicting} and Crabnet (the Compositionally Restricted Attention-Based network)\cite{wang2021compositionally} to study the impact of dataset redundancy on their performance. The Roost algorithm is a machine learning approach specifically designed for materials property prediction based on the material composition. It utilizes a graph neural network framework to learn relationships between material compositions and their corresponding properties. CrabNet  is a transformer self-attention based model for composition only material property prediction. It matches or exceeds current best-practice methods on nearly all of 28 total benchmark datasets.

\subsection{Structure based material property prediction algorithms}

We evaluate two state-of-the-art structure based material property prediction algorithms including ALIGNN (Atomistic Line Graph Neural Network)\cite{choudhary2021atomistic} and DeeperGATGNN\cite{omee2022scalable} to study the impact of dataset redundancy on their performance. The ALIGNN model addresses a major limitation of the majority of current Graph Neural Network (GNN) models used for atomistic predictions, which only rely on atomic distances while overlooking the bond angles. Actually bond angles play a crucial role in distinguishing various atomic structures and small deviations in bond angles can significantly impact several material properties. ALIGNN is a GNN architecture that conducts message passing on both the interatomic bond graph and its corresponding line graph specifically designed for bond angles. It has achieved state-of-art performances in most benchmark problems of the matbench \cite{dunn2020benchmarking}. The DeeperGATGNN algorithm is a global attention based graph neural network that uses differentiable group normalization and residual connection to achieve high performance deep graph neural networks without performance degradation. It has achieved superior results as shown in a set of material property predictions.

\subsection{Evaluation criteria}

We use the following performance metrics for evaluating dataset redundancy's impact on model performance, including Mean Absolute Error (MAE), R-squared ($R^2$), and Root Mean Squared Error (RMSE) 
 Mean Absolute Error (MAE):

\begin{equation}
\text{MAE} = \frac{1}{n} \sum_{i=1}^{n} \left| y_i - \hat{y}_i \right|    
\end{equation}

R-squared ($R^2$):

\[
R^2 = 1 - \frac{\sum_{i=1}^{n} (y_i - \hat{y}_i)^2}{\sum_{i=1}^{n} (y_i - \bar{y})^2}
\]

Where \(y_i\) represents the observed or true values, \(\hat{y}_i\) represents the predicted values, and \(\bar{y}\) represents the mean of the observed values. The summation symbol \(\sum\) is used to calculate the sum of values, and \(n\) represents the number of data points in the dataset.

\section{Results}
\label{sec:others}

\subsection{Datasets generation}

We downloaded 125,619 cif strutures from the Material Project database, which contains 89,354 unique compositions. For compositions that correspond to multiple polymorphs, we choose the average material property values as the default property value for that composition except for formation energy we use the minimum value. We also dropped the mp-101974 (HeSiO2) which has issue to calculate their Magpie features. We then remove all formulas with more than 50 atoms and got a non-duplicate composition dataset with 86,741 samples. We then use different similarity (distance) thresholds to generate non-redundant data sets. For mendeleev similarity, we use distance thresholds of 0.5, 0.8, 1, 1.5, 2, 2.5 and 3 to generate seven non-redundant datasets. The dataset sizes range from 86740 to 3177. Similarly we generate eight matsholar non-redundant datasets. The percentages of total range from 50.82\% to 2.33\%. We also applied the MD-HIT-structure algorithm to all the 125,619 cif structures and use different thresholds to generate seven XRD non-redundant datasets and eight OFM non-redundant datasets. 
After removal of redundancy based on varying degree of sample identity using MD-HIT algorithms, the details of all non-redundant datasets are shown in Table 2. 

\begin{figure}[ht!] 
    \begin{subfigure}[t]{0.33\textwidth}
        \includegraphics[width=0.99\textwidth]{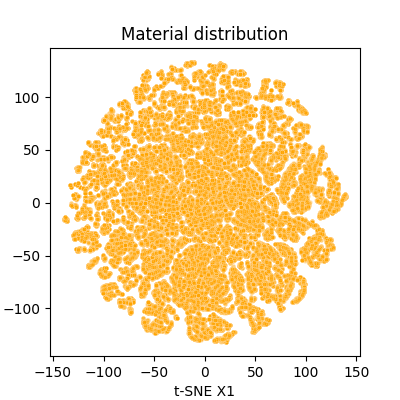}
        \caption{Whole}
        \vspace{-3pt}
        \label{fig:GaB3N4_predict2}
    \end{subfigure}
    \begin{subfigure}[t]{0.33\textwidth}
        \includegraphics[width=0.99\textwidth]{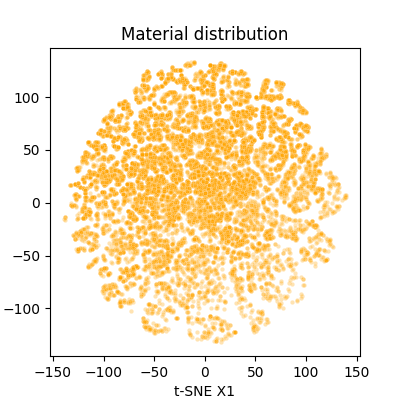}
        \caption{Matscholar-nr}
        \vspace{-3pt}
        \label{fig:GaB2N3_predict1}
    \end{subfigure} 
 \begin{subfigure}[t]{0.33\textwidth}
        \includegraphics[width=0.99\textwidth]{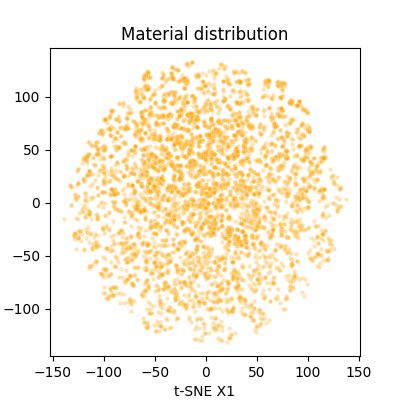}
        \caption{Mendeleev-nr}
        \vspace{-3pt}
        \label{fig:GaB2N3_target}
    \end{subfigure}              

   \caption{Distribution of whole and non-redundant MP composition datasets. (a) whole dataset with 86,740 samples. (b) non-redundant dataset using Matscholar distance with 44,081 samples. (c) Non-redundant dataset with 4,930 samples using Mendeleev distance. All maps are generated using t-SNE with Magpie composition descriptors}
  \label{fig:distribution}
\end{figure}

To visually understand the effect of redundancy removal of datasets, Figure \ref{fig:distribution} shows the material distribution t-SNE maps of the full dataset and two non-redundant datasets. For each dataset, we calculate the magpie composition features for all its samples. Then we use t-SNE dimension reduction algorithm to map the features to two dimension space. Figure 2(a) shows the distribution of whole dataset, which are filled crowded samples with high redundancy. Figure 2(b) shows the less redundant dataset Matscholar-nr generated with threshold of 0.1. It contains only 50.82\% samples while still covering the whole map. Figure 2(c) shows the Mendeleev-nr non-redundant dataset with only 4,930 samples, only 5.68\% of the whole dataset while still covering the whole map with much lower redundancy. The non-redundant datasets thus allow us to test the true generalization capability when trained and tested on them.

\begin{table}[th]
\begin{center}
\caption{Generation of non-redundant datasets }
\label{tab:datasets}
\begin{tabular}{|cccccc|}
\hline
\multicolumn{3}{|c|}{\cellcolor[HTML]{C9D9D4}Mendeleev-nr}                                                     & \multicolumn{3}{c|}{\cellcolor[HTML]{C9D9D4}Matscholar-nr}                                                    \\ \hline
\multicolumn{1}{|l}{Threshold} & \multicolumn{1}{l}{Percentage of Total} & \multicolumn{1}{l}{Dataset size} & \multicolumn{1}{l}{Threshold} & \multicolumn{1}{l}{Percentage of Total} & \multicolumn{1}{l|}{Dataset size} \\ \hline
0                              & 100.00\%                                 & 86740                             & 0                              & 100.00\%                                 & 86740                             \\
0.5                            & 46.74\%                                  & 40544                             & 0.1                            & 50.82\%                                  & 44081                             \\
0.8                            & 32.23\%                                  & 27958                             & 0.12                           & 42.56\%                                  & 36917                             \\
1                              & 24.52\%                                  & 21268                             & 0.15                           & 32.31\%                                  & 28022                             \\
1.5                            & 14.65\%                                  & 12706                             & 0.2                            & 17.86\%                                  & 15494                             \\
2                              & 8.81\%                                   & 7643                              & 0.25                           & 9.76\%                                   & 8462                              \\
2.5                            & 5.68\%                                   & 4930                              & 0.3                            & 5.50\%                                   & 4775                              \\
3                              & 3.66\%                                   & 3177                              & 0.35                           & 3.60\%                                   & 3124                              \\
\multicolumn{1}{|l}{}           & \multicolumn{1}{l}{}                     & \multicolumn{1}{l}{}              & 0.4                            & 2.33\%                                   & 2020                              \\ \hline
\multicolumn{3}{|c|}{\cellcolor[HTML]{C9D9D4}XRD-nr}                                                           & \multicolumn{3}{c|}{\cellcolor[HTML]{C9D9D4}OFM-nr}                                                           \\ \hline
\multicolumn{1}{|l}{Threshold} & \multicolumn{1}{l}{Percentage of Total} & \multicolumn{1}{l}{Dataset size} & \multicolumn{1}{l}{Threshold} & \multicolumn{1}{l}{Percentage of Total} & \multicolumn{1}{l|}{Dataset size} \\ \hline
0                              & 100.00\%                                 & 123108                             & 0                              & 100.00\%                                 & 123108                             \\
0.5                            & 50.65\%                                  & 62350                             & 0.15                            & 46.45\%                                  & 57183                             \\
0.6                            & 37.12\%                                  & 45703                             & 0.2                           & 39.32\%                                  & 48409                             \\
0.8                              & 16.98\%                                  & 20901                             & 0.45                           & 18.48\%                                  & 22748                             \\
0.9                            & 11.15\%                                  & 13729                             & 0.7                            & 10.66\%                                  & 13120                             \\
\hline
\end{tabular}
\end{center}
\end{table}

\FloatBarrier

\subsection{Composition based material property prediction with redundancy control}

\begin{figure*}[ht] 
    \centering
    \begin{minipage}[c]{0.46\textwidth}
        \centering
        \includegraphics[width=\textwidth]{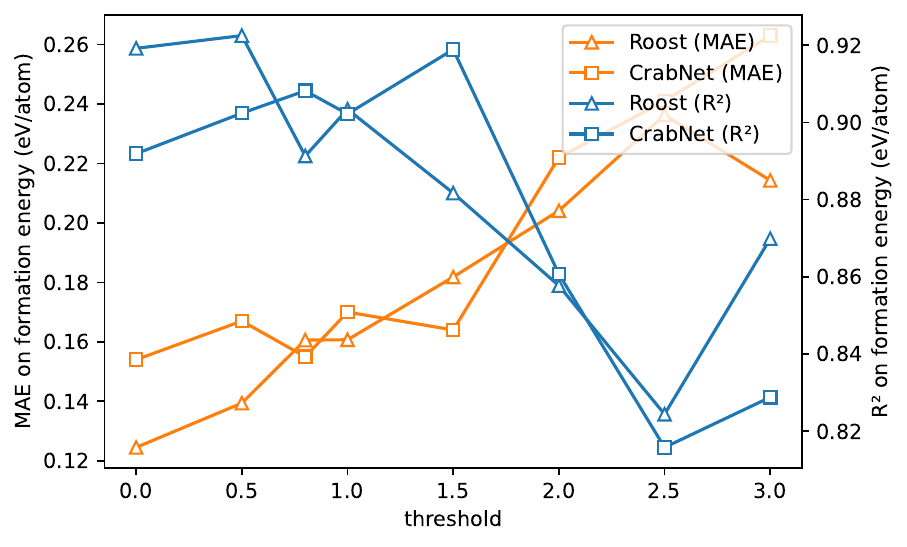}
        \subcaption{formation energy prediction}
    \end{minipage}
    \begin{minipage}[c]{0.46\textwidth}
        \centering
        \includegraphics[width=\textwidth]{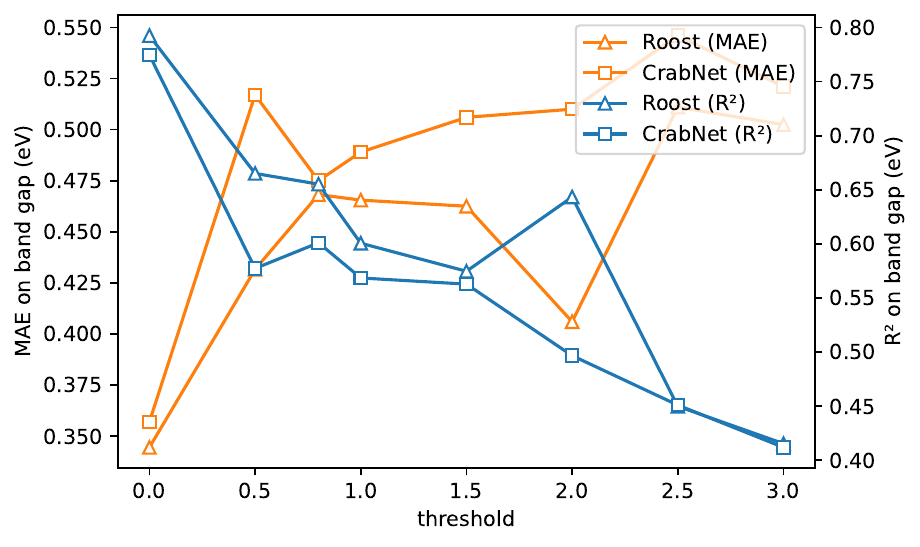}
        \subcaption{band gap prediction}
    \end{minipage}\\
    
    \caption{Performance of ML models with material properties using Mendeleev distance controlled dataset redundancy. (a) The $R^2$ (blue lines) and MAE (orange lines) results of two models trained on filtered formation energy-targeted datasets using thresholds 0.5, 1.0, 1.5, 2.0, 2.5, and 3.0. (b) The $R^2$ (blue lines) and MAE (orange lines) results of two models trained on filtered band gap-targeted datasets using thresholds 0.5, 1.0, 1.5, 2.0, 2.5, and 3.0. }
    \label{fig:mendeleev}
    
\end{figure*}

\begin{figure}[ht]
   \centering
    \begin{minipage}[c]{0.46\textwidth}
        \centering
        \includegraphics[width=\textwidth]{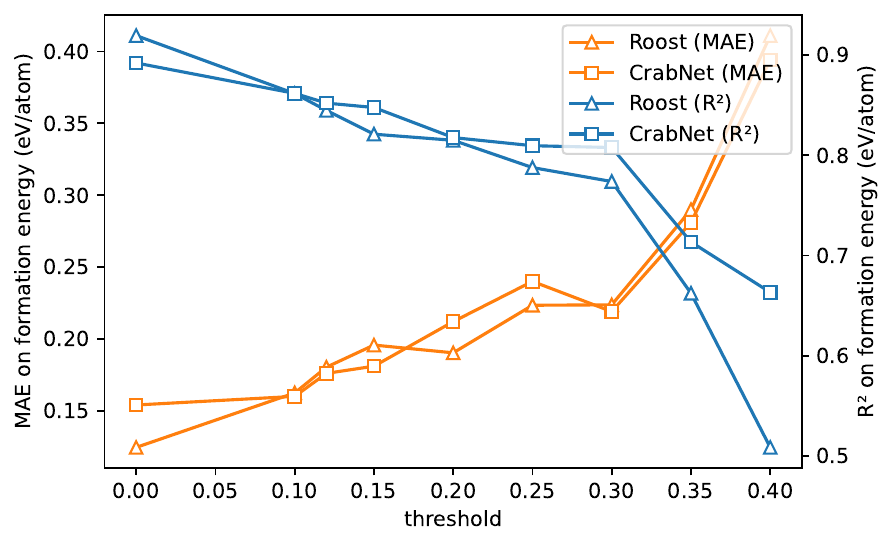}
        \subcaption{formation energy prediction}
    \end{minipage}
    \begin{minipage}[c]{0.46\textwidth}
        \centering
        \includegraphics[width=\textwidth]{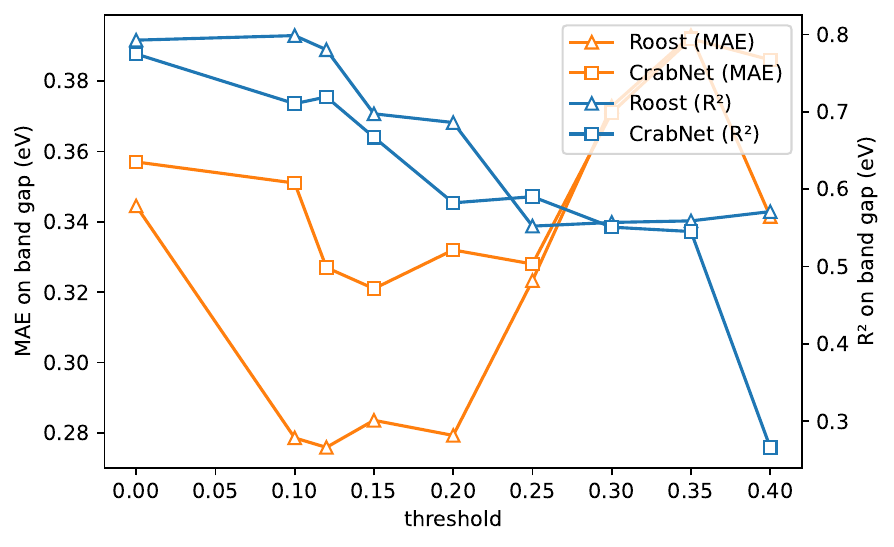}
        \subcaption{band gap prediction}
    \end{minipage}\\
    
  \caption{Performance of ML models with material properties using Matscholar distance controlled dataset redundancy. (a) The $R^2$ (blue lines) and MAE (orange lines) results of two models trained on filtered formation energy-targeted datasets using thresholds 0.05, 0.1, 0.15, 0.2, 0.25, 0.3, 0.35, 0.4. (b) The $R^2$ (blue lines) and MAE (orange lines) results of two models trained on filtered band gap-targeted datasets using thresholds 0.05, 0.1, 0.15, 0.2, 0.25, 0.3, 0.35, 0.4.}
  \label{fig:matscholar}
\end{figure}

To examine the material properties prediction performance of ML models using datasets with Mendeleev distance and Matscholar distance based redundancy control, we conducted a series of experiments to explore how the degree of redundancy affects the ML performance for formation energy and band gap prediction. The non-redundant datasets derived from the whole MP composition dataset with 86,740 samples using different thresholds were divided into training, validation, and testing sets with a ratio of 8:1:1, respectively. Figure \ref{fig:mendeleev} and \ref{fig:matscholar} show a comparison of the performances of Roost and CrabNet for formation energy and band gap prediction on datasets of different sizes, filtered by Mendeleev distance thresholds of 0, 0.5, 0.8, 1, 1.5, 2, 2.5 and 3 and Matscholar distance thresholds of 0.05, 0.1, 0.15, 0.2, 0.25, 0.3, 0.35 and 0.4. 

Figure \ref{fig:mendeleev}(a) shows the prediction performances (MAE and $R^2$) of Roost and CrabNet for formation energy prediction evaluated over the whole dataset and six non-redundant datasets. It is found that the performance of both models exhibits a deteriorating trend with the increasing thresholds corresponding to lower degree of data redundancy, as evidenced by the diminishing R2 and increasing MAE scores. For band gap prediction (Figure \ref{fig:mendeleev}(b)), the R2 scores of both models are decreasing gradually with the increase of the threshold. While the MAE scores exhibit a general uptrend, they do not exhibit a consistent decline with respect to the increasing threshold. Instead, they exhibit abrupt jumps at certain points. This could be due to outliers in the band gap-target datasets, which also shows the higher challenges for band gap prediction.

Figure \ref{fig:matscholar} shows the ML performances over the matscholar-controlled non-redundant datasets. In Figure \ref{fig:matscholar} (a), we found that the correlations between prediction performances of both Roost and CrabNet and thresholds (or data redundancy) are much higher than those shown in Figure \ref{fig:mendeleev}(a), indicating that the matscholar distance tends to generate more evenly distributed non-redundant datasets compared to Mendeleev distance. However, this consistent trends of MAE and $R^2$ do not hold in the bandgap prediction performance shown in Figure \ref{fig:matscholar}(b), in which the $R^2$ curves are similar to those found in Figure \ref{fig:mendeleev}(b) while the band gap prediction performances have large variation across different thresholds. We have checked this phenomenon by running multiple experiments for each threshold and got similar results. One possible reason is that a large percentage of bandgap samples have zero values. Overall, we found that removing redundancy of the datasets allows us to obtain more objective performances of ML models. 
Through experiments, we observe that without reducing redundancy, a significant portion of test samples are concentrated in crowded areas with low prediction errors. This occurs because the model may overly rely on the information from these redundant samples during the learning process, while disregarding other more diverse data features. Excessive sample redundancy can potentially lead to deceptive phenomena on the test set.

\FloatBarrier

\subsection{Structure based material property prediction with redundancy control}
\begin{figure}[ht]
  \centering
      \begin{minipage}[c]{0.46\textwidth}
        \centering
        \includegraphics[width=\textwidth]{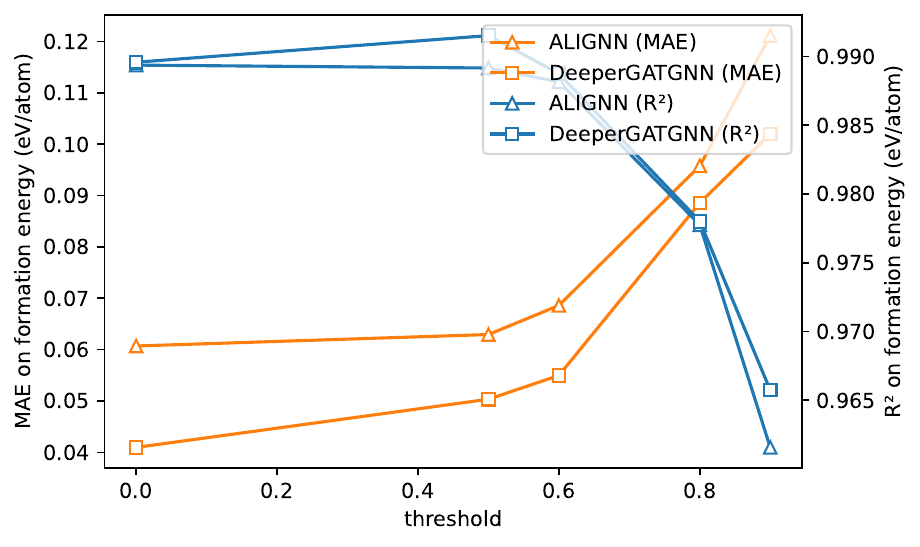}
        \subcaption{formation energy prediction}
    \end{minipage}
    \begin{minipage}[c]{0.46\textwidth}
        \centering
        \includegraphics[width=\textwidth]{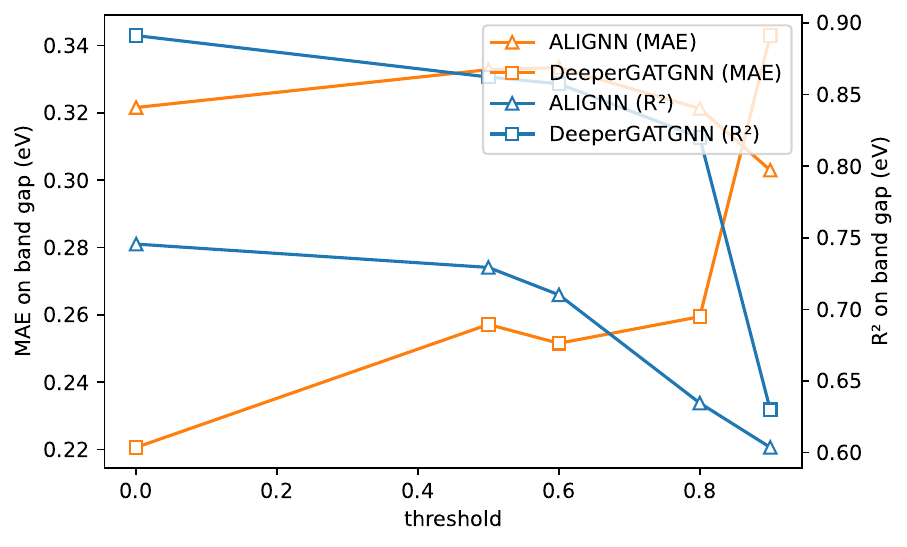}
        \subcaption{band gap prediction}
    \end{minipage}\\

  \caption{Property prediction performances of ML models based on XRD distance controlled dataset redundancy. (a) The $R^2$ (blue lines) and MAE (orange lines) results of two models trained on filtered formation energy-targeted datasets using thresholds 0.5, 0.6, 0.8, 0.9. (b) The $R^2$ (blue lines) and MAE (orange lines) results of two models trained on filtered band gap-targeted datasets using thresholds 0.5, 0.6, 0.8, 0.9.}
  \label{fig:XRD_result}
\end{figure}

To investigate the redundancy control of structure-based material datasets, we downloaded the whole Material Project database of 123,108 crystal structures along with their formation energy per atom and band gaps. Then we use the XRD and OFM features of crystal structures to define the similarity between pairs of structures, which is used to control the structure redundancy using the thresholds the minimum XRD/OFM distance between any pair of samples. For XRD based non-redundant datasets, we used the thresholds of 0.5, 0.6, 0.8, and 0.9. We then evaluated the material property prediction performances of two state-of-the-art graph neural network algorithms including DeeperGATGNN and ALIGNN. The results are shown in Figure \ref{fig:XRD_result} (a) for formation energy prediction and Figure \ref{fig:XRD_result} (b) for band gap prediction. 

First we found that the XRD-distance provides a good control of data redundancy as the MAEs of both algorithms gradually increase with the increasing XRD thresholds, corresponding to lower dataset redundancy (Figure \ref{fig:XRD_result} (a)). Simultaneously, the $R^2$ scores decrease as the thresholds go up. For band gap prediction result in Figure \ref{fig:XRD_result} (b), the degree of dataset redundancy also affects the performance of both algorithms, though with a more complex effect compared to formation energy prediction results. First, it can be found that the $R^2$ scores of both algorithms drop down with the increasing thresholds. However, while the MAEs of the DeeperGATGNN go up overall with increasing thresholds, the MAEs of ALIGNN over the non-redundant with thresholds 0.8 and 0.9 are actually lower than the result over the dataset with threshold of 0.6 while the $R^2$ scores are lower. This discrepancy indicates for the bandgap prediction problem, there is a higher nonlinearity and the outlier band gap values may also play a role here. This phenomenon is also observed in the composition based results in Figure \ref{fig:mendeleev} and Figure \ref{fig:matscholar}.

\begin{figure}[ht]
  \centering
      \begin{minipage}[c]{0.46\textwidth}
        \centering
        \includegraphics[width=\textwidth]{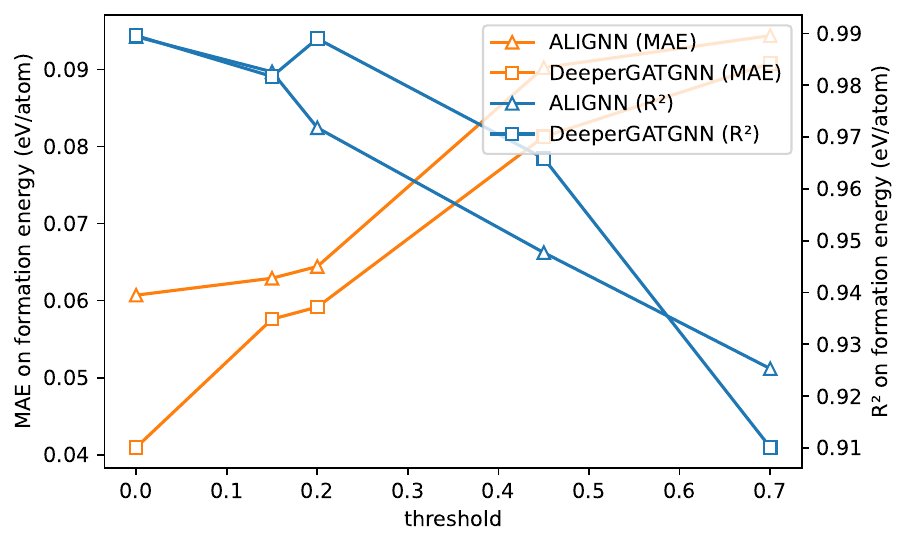}
        \subcaption{formation energy prediction}
    \end{minipage}
    \begin{minipage}[c]{0.46\textwidth}
        \centering
        \includegraphics[width=\textwidth]{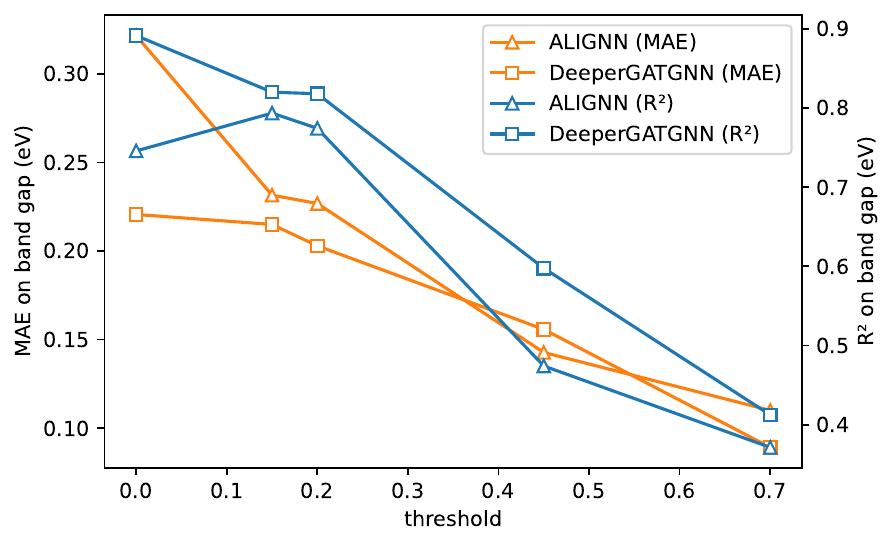}
        \subcaption{band gap prediction}
    \end{minipage}\\

  \caption{Property prediction performances of ML models based on OFM distance controlled dataset redundancy. (a) The $R^2$ (blue lines) and MAE (orange lines) results of two models trained on filtered formation energy-targeted datasets using thresholds 0.15, 0.2, 0.45, 0.7. (b) The $R^2$ (blue lines) and MAE (orange lines) results of two models trained on filtered band gap-targeted datasets using OFM thresholds 0.15, 0.2, 0.45, 0.7.}
  \label{fig:OFM_result}
\end{figure}

We further evaluated how OFM-controlled data redundancy affects the algorithms' performance. Figure \ref{fig:OFM_result}(a) and (b) show how the performances in terms of MAE and $R^2$ change with the decreasing redundancy (or increasing thresholds). First we found that both algorithms showed high consistency in the formation energy prediction (Figure \ref{fig:OFM_result}(a)). For both algorithms, the $R^2$ scores decreases in general with the increasing thresholds while the MAE scores increase. This indicates that OFM distance metric can be used as a good redundancy control method for crystal structure dataset. However, for band gap prediction, Figure \ref{fig:OFM_result}(b) shows a surprising result: the $R^2$ scores go down with the increasing threshold as expected for both algorithms. However,the MAE scores also go down, which is unexpected since lower redundancy should lead to higher challenge for property prediction. To investigate the issue, we count the percentages of near-zero bandgap (<0.01 eV) samples of the test sets for all the five datasets with thresholds 0, 0.15, 0.2, 0.45, 0.7 and found that while the whole redundant dataset contains only 48.64\% near-zero bandgap samples, our MD\_HIT algorithm accidentally tend to pick higher percentage of near-zero bandgap samples with 64.09\%, 67.81\%, 84.52\%, and 92.43\% for thresholds 0.15, 0 2, 0.45, 0.7 respectively, which makes the prediction to be much easier, which explains why the MAEs drop. To further illustrate this data bias, we plotted the scatter plots of the predicted bandgaps by DeeperGATGNN over the whole datasets and two non-redundant datasets. We can clearly see that the dominance (92.43\%) of near-zero samples in non-redundant dataset with threshold 0.7, which makes the prediction to be much easier compared to the whole dataset. This data bias may be reduced by choosing a different seed structure rather than the SrTiO$_3$ as used in this experiment. It also shows the importance to watch for data bias which can easily lead to over-estimated ML model performance in material property prediction.

\begin{figure}[ht]
  \centering
      \begin{minipage}[c]{0.33\textwidth}
        \centering
        \includegraphics[width=\textwidth]{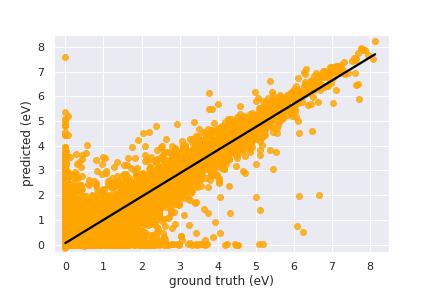}
        \subcaption{whole dataset}
    \end{minipage}
    \begin{minipage}[c]{0.33\textwidth}
        \centering
        \includegraphics[width=\textwidth]{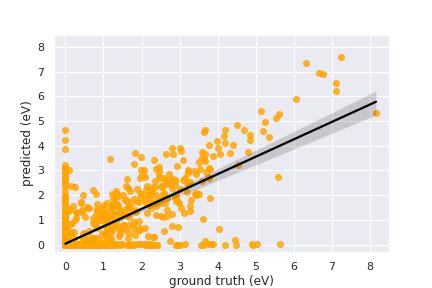}
        \subcaption{threshold 0.45}
    \end{minipage}
    \begin{minipage}[c]{0.33\textwidth}
        \centering
        \includegraphics[width=\textwidth]{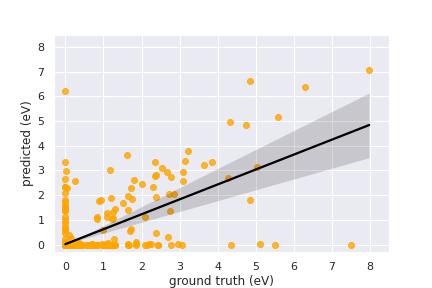}
        \subcaption{threshold 0.7}
    \end{minipage}\\
  \caption{Band gap distributions of the whole dataset and two non-redundant datasets. (a) whole dataset (b) non-redundant dataset with OFM thrshold 0.45; (c)non-redundant dataset with OFM thrshold 0.7. }
  \label{fig:bandgap_dist}
\end{figure}

\FloatBarrier

\section{Conclusion}
Large material databases such as Materials Project usually contain high degree of redundancy, which causes biased ML models and over-estimated performance evaluations due to the redundancy between randomly selected test samples and the remaining training samples. The claimed DFT accuracy averaged over all data samples from literature deviates from the common needs of material scientists who usually want to discover new materials that are different from the known training samples, which makes it important to evaluate and report the extrapolation rather than interpolation material property prediction performance.
Here we propose and develop two material dataset redundancy reducing algorithms based on a greedy algorithm inspired by the peer bioinformatics CD-HIT algorithm. We use two composition distance metrics and two structure distance metrics as the thresholds to control sample redundancy of our composition and structure datasets. Our benchmark results over two composition based and two structure based material property prediction algorithms over two material properties (formation energy and band gap) showed that the prediction performance of current ML models all tend to degrade due to the removal of redundant samples, leading to more realistic measure of prediction performance of current ML material property models. The availability of our easy-to-use open-source code of MD-HIT-composition and MD-HIT-structure makes it easy for researchers to conduct objective evaluation and report realistic peformance of their ML models for material property prediction. It should be also noted that the current multi-threaded implementation of our MD-hit algorithms are still slow and more improvements are highly desirable.

\section{Data and Code Availability}

The source code and the non-redundant datasets can be freely accessed at https://github.com/usccolumbia/MD-HIT

\section{Contribution}
Conceptualization, J.H.; methodology,J.H. Q.L.,S.L.,E.S.,Y.Z.; software, J.H., S.S.,Y.S., S.O.; resources, J.H.; writing--original draft preparation, J.H., S.S., Y.S.,S.O.,S.L.,E.S.,Y.Z.; writing--review and editing,  J.H; visualization, J.H. and S.S.; supervision, J.H.;  funding acquisition, J.H.

\section*{Acknowledgement}
Qin Li would like to thank for the computing support of the State Key Laboratory of Public Big Data, Guizhou University.

\bibliographystyle{unsrt}  
\bibliography{references}

\end{document}